\documentclass{article}
\usepackage{amssymb}
\begin{document}

\title{Comments on regularization ambiguities and local gauge symmetries}
\author{R. Casana\thanks{casana@ift.unesp.br} and B. M. Pimentel\thanks{
pimentel@ift.unesp.br} \\
{\small Instituto de F\'{\i}sica Te\'orica, Universidade Estadual
Paulista} \\ {\small Rua Pamplona 145, CEP 01405-900, S\~ao Paulo, SP,
Brazil}}

\date{}

\maketitle

\begin{abstract}
We study the regularization ambiguities in an exact renormalized
(1+1)-dimensional field theory. We show a relation between the
regularization ambiguities and the coupling parameters of the theory
as well as their role in the implementation of a local gauge
symmetry at quantum level.
\end{abstract}


{\small PACs: 03.70.+k, 11.15.-q, 11.30.-j}

\section{Introduction}

The study of anomalies in local gauge theories has been very important for
the definition of the modern Standard model of elementary particles. The
search for local gauge theories without gauge anomalies established the
existence of equal number of families of leptons and quarks \cite{sm-1}.
But, in the Standard model assumes the existence of a fundamental spin 0
particle, the Higgs boson \cite{sm-2}, it allows to generate, in a gauge
invariant way, mass for the vectors bosons of the Electroweak sector via
spontaneous symmetry breaking \cite{weinberg}. This procedure does not spoil
the perturbative renormalizability of the model \cite{sm-3}. Nevertheless,
to date there is no experimental evidence about the existence of this
particle.

On the other hand, there is an alternative mechanism to generate
mass, the so called dynamical mass generation. As it was shown first
by Schwinger \cite{schwinger} (and at the first time it was
suggested by Yang and Mills in their seminal paper \cite{y-m}), it
can generate mass for the vector boson at quantum level in a gauge
invariant way. But, the dynamical mass generation can also be
carried in a non-gauge invariant way, just as it happens, for
example, in the chiral Schwinger model (CSM)
\cite{jackiw1,Jackiwtop} or in the anomalous Schwinger model (ASM)
\cite{mitra1,trab1}.

In quantum field theory is natural the arising of divergences when
computing, for example, the correlation functions of a specific model. The
calculus involved, sometimes, carries on to the rising of ambiguities linked
to the prescription used to regularize these infinities \cite%
{jackiw1,Jackiwtop,jackiw0}.

In this paper we mixed both the dynamical mass generation and the
regularization ambiguities such that a non-gauge invariant model at
classical level can gain a local gauge symmetry at quantum level. In
order to show it, we analyze a simple (1+1) dimensional model: a
massive vector (Proca) field coupled to massless Dirac's fermions
known as the Thirring-Wess model \cite{thirring-wess}. As it is
well-known this model does not have a local gauge symmetry at
classical level, then, we compute the fermionic determinant in a
external field with a generalized prescription which does not
preserve the fermionic current, i. e., $\partial_\mu \langle 0 |
\overline\psi \gamma^\mu \psi|0\rangle_{\!A_\mu} \neq 0$ \cite{Das}.
The generalized prescription introduces an ambiguity regularization
parameter such as it happens in the CSM or ASM models. The fermionic
determinant contributes with an additional mass for the vector field
which is ambiguity dependent. Thus, the effective action depending
in the value assumed for the ambiguity can gain or not a local $U(1)$ symmetry (%
$A_\mu \rightarrow A_\mu+\partial_\mu\alpha$). Therefore, when we impose the
local gauge symmetry in the effective action level we get to define the
ambiguity parameter as a full function of the coupling constants, and the
model is a well-known finite field theory. On the other hand, if we do not
impose such local symmetry the model get still to be a consistent and
renormalizable field theory.

Consequently, we can assert that this non--gauge invariant classical model
suggest the possibility of generating a local gauge invariance by quantum
effects.

The paper is displayed in the following way. In section 2, we
quantize and analyze the massive vector field coupled to massless
Dirac's fermions in both situations, and the important results are
analyzed in section 3.


\section{The model}


The Thirring-Wess model \cite{thirring-wess}  is defined by the following
Lagrangian density\footnote{%
In this paper, we use the natural units $\hbar =c=1$ and the following
conventions, the metric tensor is $g_{00}=1=-g_{11}$, the Levi-Civita tensor
is $\epsilon ^{01}=1=-\epsilon _{01}$. The $\gamma _{\mu }$ matrices satisfy
$\{\gamma ^{\mu },\gamma ^{\nu }\}=2\,g^{\mu \nu }$, the $\gamma _{5}$
matrix is defined as $\{\gamma _{\mu },\gamma _{5}\}=0\;,\;\;\gamma
_{5}=\gamma ^{0}\gamma ^{1}$, and the chiral projectors are $P_{\pm }=\frac{1%
}{2}(1\pm \gamma _{5})$. Also define the contraction $\tilde{V}^{\mu
}=\epsilon ^{\mu \nu }V_{\nu }$.}
\begin{equation}
\mathcal{L}[\psi ,\overline{\psi },A]=-\frac{1}{4}F_{\mu \nu }F^{\mu \nu }+%
\frac{m_{P}^{2}}{2}\;A_{\mu }A^{\mu }+\overline{\psi }(\;i\partial \!\!\!%
\slash+eA\!\!\!\slash\;)\psi ,  \label{pr1}
\end{equation}%
as it is clear, the mass term of vector field precludes the existence of
local $U(1)$ gauge symmetry at classical level.

We quantize the model by introducing the generating functional
\begin{eqnarray}  \label{pr2}
Z[\eta,\overline\eta,J^\mu]=N\!\int\! dA_\mu d\psi d\overline\psi\;
\exp\left(i \int\!\! dx \;-\frac{1}{4}F_{\mu\nu} F^{\mu\nu}+
\frac{m^2_P}{2}\;A_\mu A^\mu\,+\, \overline\psi (\;i
\partial\!\!\!\slash +eA \!\!\!\slash\;) \psi++ J_\mu A^\mu+ \overline
\eta\psi+\overline\psi\eta\right)
\end{eqnarray}
where $\eta,\overline\eta$ are the external sources for the fermionic fields
and $J^\mu$ is the correspondent source for the vector field. We consider as
the first step, in the quantization procedure, the computing of the
fermionic integration which gives
\begin{eqnarray}
\det\left(\frac{}{}i\partial\!\!\!\slash +eA \!\!\!\slash\,\right)
\;\exp\left(-i\!\!\int\!\! dx\,dy \; \overline \eta(x) G(x,y;A)
\eta(y)\right).
\end{eqnarray}
The computation of the fermionic determinant will be carry out with a
generalized prescription \cite{Jackiwtop,Das} which does not necessarily
preserve the fermionic current, and it gives the following contribution to
the effective action for the massive vector field,
\begin{eqnarray}  \label{pr3}
\det\left(\frac{}{}i\partial\!\!\!\slash +eA \!\!\!\slash\,\right) =\int\!\!
dx\;\frac{1}{2} A_\mu(x)\left(\frac{e^2}{2\pi}(a+1) \, g^{\mu\nu} -\frac{e^2%
}{\pi}\frac{\partial^\mu \partial^\nu}{\square} \,\right)A_\nu(x),
\end{eqnarray}
where we can see that the vector field get an additional mass, which
is dynamically generated. The $a$ parameter characterizes the
regularization ambiguities of the UV divergences which appear when
compute the fermionic determinant.

Thus, after integration on the fermion fields in (\ref{pr2}), it reads as
\begin{eqnarray}  \label{pr4}
Z[\eta,\overline\eta,J^\mu]&=& \!\int\!\!dA_\mu \,\exp\! \left(i\!
\!\int\!\!\! dx\frac{1}{2}A_\mu \!\!\left\{g^{\mu\nu} \!\left[\frac{}{}\!
\square\!+\!m^2\right]\!- \!\left[ \square \!+\!\frac{e^2}{\pi}\right] \frac{%
\partial^\mu\partial^\nu} {\square}\right\}\!\!A_\nu \!+\!J\cdotp{A} \right)
\\
& & \qquad\times\exp\left(-i\!\!\int\!\! dx\,dy \; \overline \eta(x)
G(x,y;A)\eta(y)\right)  \nonumber
\end{eqnarray}
where $m^2$ is the total mass of the vector field,
\begin{eqnarray}  \label{mass_vec}
m^2=m^2_P+\frac{e^2}{2\pi}(a+1) =\frac{e^2}{2\pi}(b+1)
\end{eqnarray}
and the $b$ parameter is defines as
\begin{eqnarray}  \label{pr6}
b\;=\;a+2\pi\frac{m^2_P}{e^2}\;.
\end{eqnarray}
$G(x,y;A)$ is the Green's function of the Dirac's equation,
\begin{eqnarray}  \label{pr7x}
\left(i\partial\!\!\!\slash + eA\!\!\!\slash\right)G(x,y;A)= \delta(x-y)\;,
\end{eqnarray}
it can exactly be computed,
\begin{equation}  \label{pr8}
G(x,y;A)=\exp\left(-ie\int\!\! dz\;A_\mu(z) j^\mu_+(z,x,y) \right)
P_+G_F(x-y) + \exp\left(-ie \int\!\!
dz\;A_\mu(z)j^\mu_-(z,x,y)\;\right) P_-G_F(x-y)\,
\end{equation}
with $G_F(x-y)$ being the Green's function of the free Dirac's equation: $%
i\partial \!\!\!\slash\, G_F(x-y)= \delta(x-y)\,\;$ and the contact
current $j^\mu_\pm(z,x,y)$ is
\begin{eqnarray}  \label{pr9}
j^\mu_\pm (z,x,y)=(\partial^\mu_z \mp \tilde\partial^\mu_z) [D_F(z-x) -
D_F(z-y)]\;,
\end{eqnarray}
where $D_F(x)$ is the Green's function of the massless Klein--Gordon
equation: $\;\square\, D_F(x-y)= \delta(x-y)\,$.

\vskip 0.3cm

At quantum level the Ward identity for the vectorial 1PI two-point
function is given by
\begin{eqnarray}  \label{pr7}
k_\mu \Gamma^{\mu\nu}(k)=\frac{e^2}{2\pi}(b-1)\;k^\nu\;,
\end{eqnarray}
then, we will have a local $U(1)$ gauge symmetry  if and only if
$b=1$.

Therefore, we have to analyze the remaining vector quantization in
the generating functional (\ref{pr4}) by distinguishing the case
$b=1$ (which we named as the quantum local gauge invariant model)
from $b\neq 1$ case (named as the quantum non--gauge invariant
model).


\subsection{The $b=1$ case}


In this way the possibility to get a quantum local gauge theory from
the model (\ref{pr1}) turns into a reality. We first set $b=1$ in
the generating function (\ref%
{pr4}), such procedure leads us up to an equation for the ambiguity
parameter $a$ as a function of the coupling constants of the classical field
theory,
\begin{eqnarray}  \label{pr38}
a=1-2\pi\frac{m^2_P}{e^2}\;.
\end{eqnarray}
at this level we can say that the ambiguity parameter $a$ is related to the
weak and strong coupling regime of the model, i. e., $m^2_P<e^2$ and $%
m^2_P>e^2$. And consequently, the vectorial mass (\ref{mass_vec}),
would be given for
\begin{eqnarray}  \label{pr16x}
m^2=\frac{e^2}{\pi}
\end{eqnarray}
it is well-defined. And, we recover the well-known local gauge invariant
Schwinger model \cite{schwinger}.

This way, the requirement to get a local gauge symmetry at quantum level in
the effective action for the vector field leads us to find a highly physical
meaning for the ambiguity parameter. We find a way of turning the ambiguity
parameter, which was known to be unrelated to the theory, an element of the
full quantized theory. This open a possibility to redefine an anomalous
gauge model or non--renormalizable model (in the usual sense) to becomes a
well defined and renormalizable field theory and, therefore, to be capable
to predict physical phenomena.

Now, it is worthwhile to clarify that when $b=1$ the theory is a gauge
theory, then it has to be quantized considering its local gauge symmetry. By
example, we can follow the Faddeev-Popov technique \cite{faddeev} for gauge
theories, whose finality is to lead the functional integration well--defined.


\subsection{The $b\neq 1$ case}


Now, we analyze the quantum non--gauge invariant model or $b\neq 1$
case. In this situation, the Thirring-Wess model is exactly equal,
in all aspects, at quantum level to the anomalous Schwinger model
studied in \cite{trab1,trab2,trab3}; but in this case the $a$
ambiguity parameter from the cited references is replaced by the $b$
parameter. In these references the authors made a detailed study of
the ASM model, such as semi--perturbative renormalizability
\cite{trab1}, determination of non--perturbative character of UV
divergences \cite{trab2}, also its exact renormalization by summing
the regularized semi--perturbative
expansion and the consequences on the physical properties of the model \cite%
{trab3}.

Thus, from equation (\ref{pr4}), we can compute all Green's functions of the
non-gauge invariant model. For example, the vectorial propagator is
straightforwardly computed, it yields (in momentum space) (see \cite{trab1},
for example)
\begin{equation}  \label{pr10}
i\,G_{\mu\nu}(k)= \frac{1}{k^2-m^2}\left(g_{\mu\nu} -\frac{k_\mu k_\nu}{k^2}%
\right)-\frac{2\pi}{e^2(b-1)}\; \frac{ k_\mu k_\nu}{k^2},
\end{equation}
the transversal part of vector field has a modified mass which is
\begin{eqnarray}  \label{pr11}
m^2=m^2_P+\frac{e^2}{2\pi}(a+1)=\frac{e^2}{2\pi}(b+1)\;.
\end{eqnarray}
with $b \geq -1$ to avoid tachyonic particles. As we can notice, the
mass remains indefinite due the explicit dependence in the
regularization ambiguity.

\section{Discussion and conclusion}

We have studied a simple (1+1)--dimensional field theory which does
not have a local gauge symmetry at classical level i.e. the
Thirring-Wess model. Because, it is not a local gauge model the
fermionic determinant can be computing by using a generalized
prescription to regularize the ultraviolet divergences. This
generalized procedure leads to the appearing of regularization
ambiguities during the quantization process (read as computation of
the functional integration) characterized by the $a$ parameter.
Thus, upon the fermionic integration, the effective action for the
Proca field can would get a local
gauge symmetry if we choose an special value for the ambiguity $a$ (\ref%
{pr38}). The resulting gauge theory is the well--known Schwinger
model \cite{schwinger}.

On the other hand, if we left the ambiguity parameter arbitrary the
effective action is non local gauge invariant, and the resulting theory is
exactly equivalent to the anomalous Schwinger model studied in \cite%
{mitra1,trab1,trab2,trab3}.

The possibility of exploring the ambiguity parameter and get a local
symmetry at quantum level it was also observed in the massless
Thirring model \cite{thirring1}, where such requirement allows to
found a new phase to the model. Thus, it leads us to fix the
ambiguity parameter as a explicit function of the coupling constants
of the theory.

The model studied has the possibility to be extended to higher
dimensions but in this case the results will be only of perturbative
character. Therefore, it can would be interesting to know, at least,
the 1-loop properties of the (1+2)-D model.


\subsection*{Acknowledgements}
R.C. thanks to  FAPESP (grant 01/12611-7) for full support and
B.M.P. thanks CNPq and FAPESP (grant 02/00222-9) for partial support


\end{document}